\documentclass[aps,prl,reprint,superscriptaddress,twocolumn]{revtex4-2}
\usepackage{graphicx}
\usepackage{siunitx}
\usepackage{xcolor}
\usepackage[normalem]{ulem}
\usepackage{amsmath}
\begin{document}

\title{Spectroscopic evidence for a new type of surface resonance at noble metal surfaces}

\author{Tobias Eul}
\email[]{teul@rhrk.uni-kl.de}
\affiliation{Technische Universität Kaiserslautern and Research Center OPTIMAS, 67663 Kaiserslautern, Germany}
\author{Jürgen Braun}
\affiliation{Department Chemie, Ludwig-Maximilians-Universität München, 81377 München, Germany}
\author{Benjamin Stadtmüller}
\affiliation{Technische Universität Kaiserslautern and Research Center OPTIMAS, 67663 Kaiserslautern, Germany}
\author{Hubert Ebert}
\affiliation{Department Chemie, Ludwig-Maximilians-Universität München, 81377 München, Germany}
\author{Martin Aeschlimann}
\affiliation{Technische Universität Kaiserslautern and Research Center OPTIMAS, 67663 Kaiserslautern, Germany}

\date{\today}

\begin{abstract}
We investigate the surface- and bulk-like properties of the pristine (110)-surface of silver using threshold photoemission by excitation with light of \SI{5.9}{\electronvolt}.
Using a momentum microscope, we identified two distinct transitions along the $\overline{\Gamma}\,\overline{\textrm{Y}}$-direction of the crystal.
The first one is a so far unknown surface resonance for the (110) noble metal surface, exhibiting an exceptionally large bulk character, that has so far been elusive in surface sensitive experiments.
The second one stems from the well known bulk-like Mahan cone oriented along the $\Gamma L$-direction inside the crystal but projected onto the (110)-surface cut.
The existence of the new state is confirmed by photocurrent calculations and its character analyzed.
\end{abstract}

\maketitle

In solid-state systems, all electronic and optical properties are intrinsically linked to the band structure of the material.
This fundamental connection has started the quest to discover novel electronic states in conventional or exotic materials and to uncover their role for the material properties.
As a result, it is nowadays widely accepted that most properties of solids can be attributed to the dispersion and the orbital character of either \textit{bulk bands} or \textit{surface states}\cite{Speer.2009}.
For instance, bulk bands are responsible for the optical and transport properties of materials while surface states dominate the surface reactivity\cite{Forster.2008} or the catalytic properties of the surfaces of bulk materials. 

\begin{figure}[!b]
\includegraphics{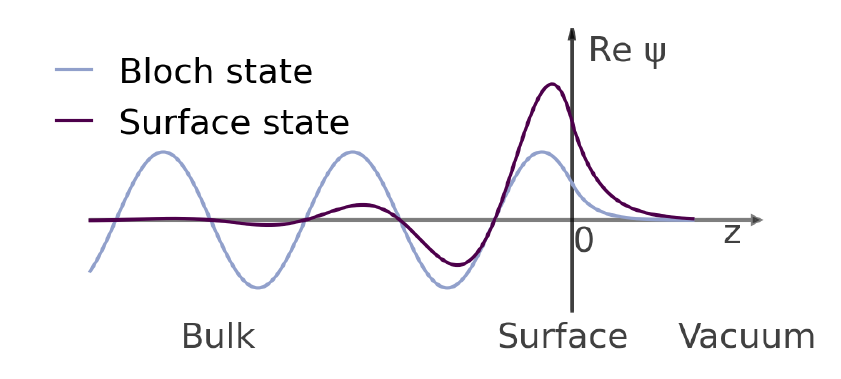}%
\caption{Real part of the wave functions of Bloch and surface states. The surface state is strongly localized in the first layers, whereas the Bloch wave propagates deep into the bulk. Both wave functions decay exponentially above the surface.}
\label{fig:figure1}
\end{figure}

From a fundamental point of view, the distinct separation between bulk and surface states is intrinsically rooted in their different spatial expansion perpendicular to the surface. 
Bulk electrons strictly follow the transition symmetry of the crystal in three dimensions and are hence described by Bloch states as shown in Fig.~\ref{fig:figure1}.
These Bloch states are characterized by the crystal momentum of the electrons parallel $\vec{k}_{\parallel}$ and perpendicular $\vec{k}_{\perp}$ to the surface and are invariant with respect to the translation of the crystal momentum $\vec{k}$ by any vector $\vec{G}$ of the reciprocal crystal lattice. 
Accordingly, the energy spectrum of bulk states, i.e., the band structure, is described by $\vec{k}_{\parallel}$ and $\vec{k}_{\perp}$ and hence reveals a strong band dispersion perpendicular to the surface. 
In contrast, the wave functions of surface states are strongly localized in the first surface layers (see Fig.~\ref{fig:figure1}) and their wave function and band structure only depend on the momentum of the electrons $\vec{k}_{\parallel}$  parallel to the surface.
This is most prominently known for the Shockley surface states of noble metal surfaces\cite{Gartland.1975,Nicolay.2000,Unal.2011,Tamai.2013,Yan.2015,Yaji.2018} or surface states of topological insulators\cite{Fu.2007,Fu.2007b,Kane.2008,Xia.2009,Chen.2009,Hasan.2010,Jozwiak.2013,Seibel.2015}.

In a vast variety of materials, the electronic surface band structure is even richer and exhibits so-called \textit{surface resonances}.
They are hybrids between surface and bulk states with different surface and bulk character depending on the hybridization strength between their constituents\cite{HansLuth.2010}.
So far however, only few studies focused on surface resonances, mainly in metals\cite{Braun.2014}, topological insulators\cite{Jozwiak.2016,PhysRevB.95.125405} and black phosphorus\cite{Golias.2016,Ehlen.2018}.
In these cases, the observed surface resonances are largely dominated by the contributions of the surface states.

In this letter, we revisit the surface band structure of a pristine (110)-surface of a silver single crystal, which has been of interest in other recent photoemission studies\cite{AndiLi.2020}.
For this surface, a variety of different surface states and surface resonances with dominant surface character have already been reported\cite{Ho.1980,Reihl.1984,A.Goldmann.1985,Bartynski.1986,Altmann.1986,L.E.Urbach.1992,Knoesel.1996b,A.Gerlach.1999,Pascual.2001}.
The band dispersion of these states is independent on the electron momentum $k_\perp$  and they exhibit predominant $p_z$-like orbital character.
Both properties are reflected in unique spectroscopic signatures of such crystal-derived surface states, namely the photon energy independent band dispersions $E(\vec{k}_\parallel)$ and significantly stronger photoemission yield for optical excitations using p-polarized light.
In our work, we provide compelling evidence for the existence of a new type of surface resonance in the silver surface band structure, which we refer to as \textit{surface scattering resonance}.
It arises from additional surface contributions to the Fourier sum of the bulk Bloch states and has a strongly light polarization dependent emission pattern as well as a clear band dispersion along $\vec{k}_{\perp}$.
A previous photoemission study\cite{AndiLi.2020} mentioned its observation, but did not disclose its origin and character.

\begin{figure}[!t]
\includegraphics[width=0.45\textwidth]{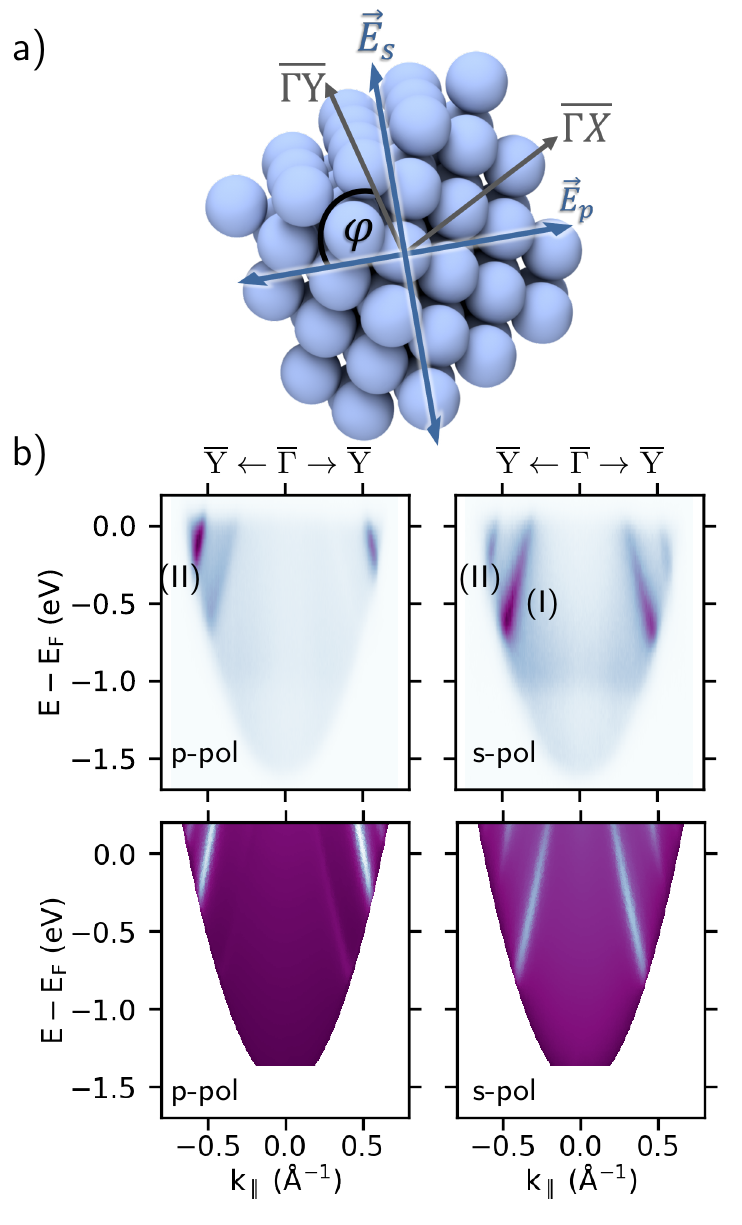}%
\caption{a) Sample orientation in the experiment. The angle $\varphi$ between the $\overline{\Gamma}\,\overline{\textrm{Y}}$-axis and the in-plane component of the p-polarized electric field vector equals \SI{60}{\degree}. b) Measured (top) and calculated (bottom) energy distribution maps cut along $\overline{\Gamma}\,\overline{\textrm{Y}}$ for p- (left) and s-polarization (right), under \SI{65}{\degree} incidence angle with a photon energy of \SI{5.9}{\electronvolt}}.
\label{fig:figure2}
\end{figure}

For the experimental part of our study, we employ a photoemission electron microscope combined with a time-of-flight delayline detector installed in an ultrahigh vacuum chamber.
The PEEM is operated in momentum space mode and records three-dimensional datasets ($k_\parallel^x, k_\parallel^y, E_{kin}$) in a single data acquisition with a resolution of $\Delta k > \SI{2e-2}{\angstrom^{-1}}$ and $\Delta E > \SI{40}{\milli\electronvolt}$.
The sample is illuminated with linearly polarized light under near normal incidence (NI, \SI{4}{\degree}) or grazing incidence (GI, \SI{65}{\degree}) with respect to the surface normal.
The orientation of the polarization of the incoming photons with respect to the high symmetry directions of the crystal surface is shown in Fig.~\ref{fig:figure2}a).
The orientation of the crystal was assessed by low-electron energy diffraction (LEED).
Here, the $\overline{\Gamma}\,\overline{\textrm{Y}}$-direction ($\overline{\Gamma}\,\overline{\textrm{X}}$-direction) corresponds to the (001)-direction ((1$\overline{1}$0)-direction) in real space, which we define as the x-axis (y-axis).
In the grazing incidence geometry (\SI{65}{\degree}), the in-plane electric field component $\vec{E}_p$ of p-polarized light is rotated by $\varphi=\SI{60}{\degree}$ with respect to the $\overline{\Gamma}\,\overline{\textrm{Y}}$-axis.
In contrast, for the normal incidence geometry, the in-plane component of the electric field can be changed to any arbitrary angle $\varphi$ using a half wave plate.

For our light source, we use \SI{5.9}{\electronvolt} photons (fourth harmonic of a Ti:Sa laser oscillator) suitable for observing one-photon photoemission at noble metal surfaces while maintaining the bulk-sensitivity of the experiment.

In conjunction with our experiment, we performed self-consistent electronic structure calculations within the {\em ab-initio} framework of density functional theory to support the interpretation of the experimental data.
For this, we use the Vosko, Wilk, and Nusair parametrization to supply the exchange and correlation potential\cite{S.H.Vosko.1980}.
We calculate the electronic structure in a fully relativistic mode by solving the corresponding Dirac equation, using the relativistic multiple-scattering or KKR formalism in the TB-KKR mode\cite{SPR-KKR8.5,Ebert.2011,Ebert.2000}.
Finally, we base the photocurrent calculations on the resulting half-space electronic structure represented by single-site scattering matrices for the different layers including the wave functions for the corresponding initial and final state energies as additional input quantities.
Most importantly, we were able to vary the bulk sensitivity of the calculated photocurrent by altering the lifetimes of the initial and final states, which effectively influences the inelastic mean free path (IMFP).
Additional details can be found in the supplemental material.

We start our discussion with an overview of near threshold angle resolved photoemission data of the Ag(110) surface.
The top row of Fig~\ref{fig:figure2}b) shows the experimental energy distribution maps (EDM) extracted along the $\overline{\Gamma}\,\overline{\textrm{Y}}$-direction for p- (left) and s-polarized light (right), respectively.
The bottom row shows the corresponding theoretical EDMs calculated for the identical experiment geometry and conditions, which agree well with the experimental data.
Minor deviations are attributed to slight differences of the theoretical and experimental Fermi energy due to a well-known shortcoming of the local density approximation (LDA)\cite{HEckardt.1984}.

For both light polarizations we detect two signatures with strong bulk character, which we refer to as features (I) and (II) as shown in Fig.~\ref{fig:figure2}b).
In the following discussion, we will provide clear evidence that these states can be assigned to a Mahan-cone-like bulk transition (I) and a new surface resonance (II). 
The Mahan-cone (I) reveals a strong intensity dependence on the polarization of the exciting light.
It is only visible for s-polarized excitation and suppressed for p-polarized excitation.
The second feature (II) can be clearly recognized for both light polarizations. 
The different polarization dependency of both transitions (I) and (II) is further reflected in the photocurrent obtained in the normal incidence geometry.
\begin{figure}[!b]
\includegraphics[width=0.45\textwidth]{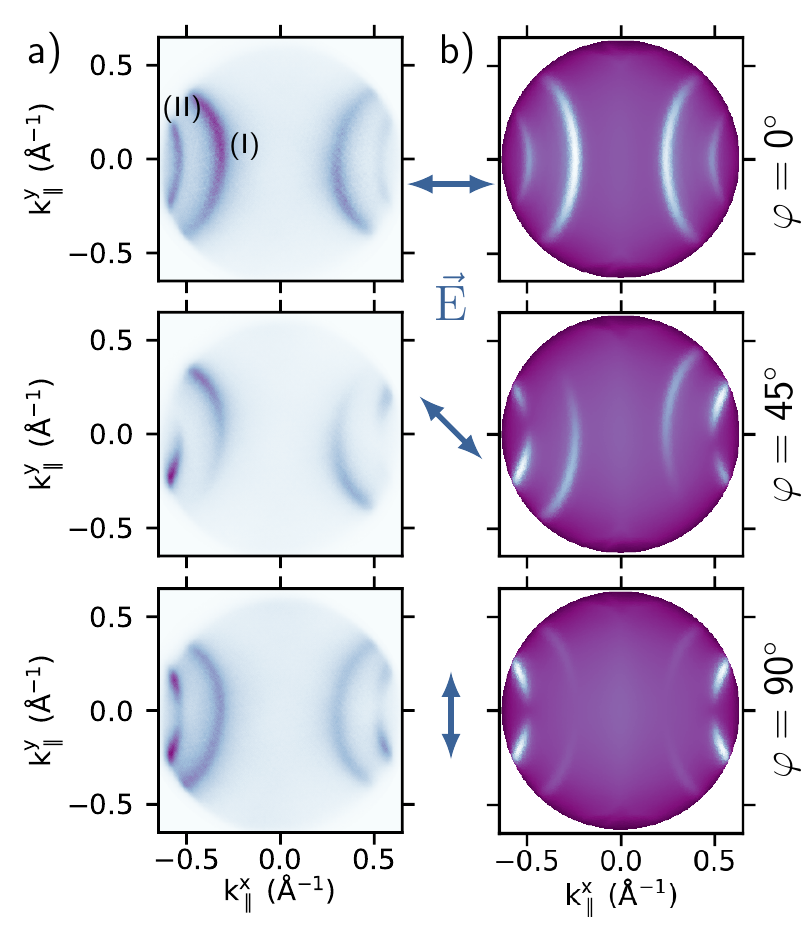}
\caption{Constant binding energy maps for photoemission in the normal incidence excitation geometry for a photon energy of \SI{5.9}{\electronvolt} with varying angle $\varphi$ between the polarization and the $\overline{\Gamma}\,\overline{\textrm{Y}}$-axis. The images were taken at $E-E_F = \SI{0}{\electronvolt}$. a) Experiment. b) Theory. The $k_\parallel^x$-axis corresponds to the $\overline{\Gamma}\,\overline{\textrm{Y}}$-direction in k-space and the x-axis in real space.}
\label{fig:figure3}
\end{figure}
Figure~\ref{fig:figure3} shows constant binding energy maps (CBMs) at the Fermi energy $E-E_F=\SI{0}{\electronvolt}$ for the angles $\varphi\in\{\SI{0}{\degree},\SI{45}{\degree},\SI{90}{\degree}\}$, with the experimental and theoretical data in the left and right column, respectively (CBMs for additional angles can be found in the supplemental material).
In the CBMs, the features (I) and (II) appear as parabolic emission patterns with their center points located on the $\overline{\Gamma}\,\overline{\textrm{Y}}$-axis.
Both features (I) and (II) show a polarization dependent photoemission pattern, characterized by inhomogeneities of the intensity for angles between the axes as evident for $\varphi=\SI{45}{\degree}$ as well as a local minima for $k_\parallel^y=\SI{0}{\angstrom^{-1}}$ in the case of $\varphi=\SI{90}{\degree}$.
For $\varphi=\SI{45}{\degree}$, the calculated intensity pattern fits quantitatively for feature (II) whereas feature (I) appears mirrored at the horizontal axis in contrast to the measurement.
Furthermore, the intensity of feature (I) decreases for $\varphi=\SI{90}{\degree}$ while being the brightest for  $\varphi=\SI{0}{\degree}$.
On the other hand feature (II) gains more visibility rotating the polarization from \SI{0}{\degree} to \SI{90}{\degree}. 
These characteristic changes of the photoemission patterns clearly point to a different microscopic origin of both features.

\begin{figure}[!t]
\includegraphics[width=0.45\textwidth]{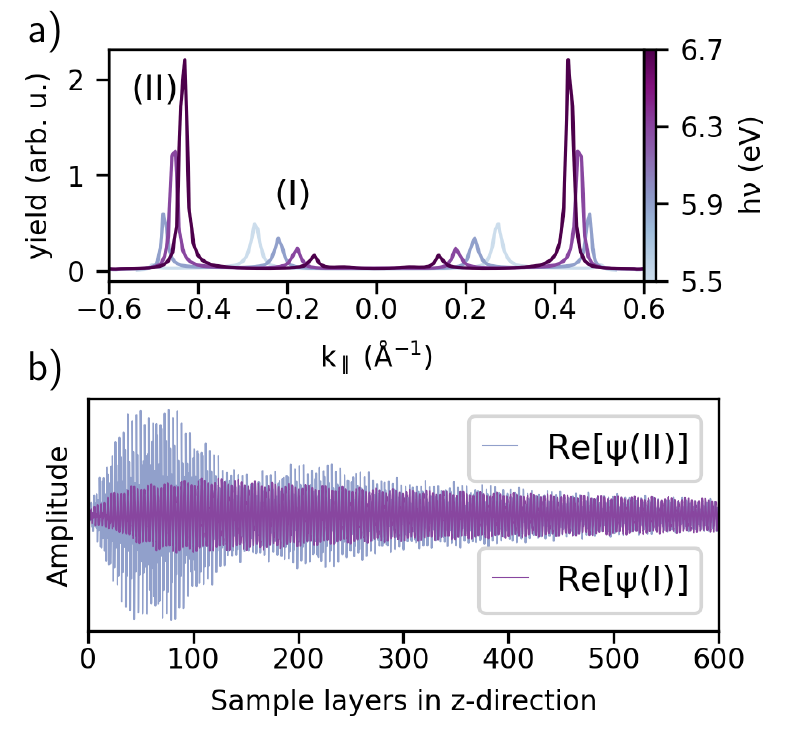}
\caption{a) Photoemission intensity for a cut along the $\overline{\Gamma}\overline{Y}$-direction for different photon energies. b) Calculated initial state wave functions for a photon energy of \SI{6.7}{\electronvolt} and a polarization angle of \SI{45}{\degree}.}
\label{fig:figure4}
\end{figure}

In the next step, we investigate the photon energy dependent photocurrent of the Ag(110) surface band structure.
This information can be extracted from our one-step photoemission simulations, which was able to describe the main aspects of our experimental photoemission data with high accuracy.
We calculated CBMs for an initial state energy of $E-E_{F}=\SI{0}{\electronvolt}$, a polarization angle of \SI{45}{\degree} and photon energies between $\SIrange{5.5}{6.7}{\electronvolt}$.
The corresponding 1D intensity profiles extracted at $k_\parallel^y=\SI{0}{\angstrom^{-1}}$ are shown in Fig.~\ref{fig:figure4}a).
The spectroscopic features (I) and (II) reveal two well-defined maxima with a photon energy dependent photoemission yield.
While the intensity of feature (II) increases with increasing photon energy, the one of feature (I) decreases and almost vanishes at a photon energy of $\SI{6.7}{\electronvolt}$.
This points to a different cross section of both features and hence again to a different microscopic origin of both states.
More importantly, the maxima of both features (I) and (II) change their position in momentum space when changing the photon energy.
This behavior is fully in line with a significant band dispersion of both features for $\vec{k}_{\perp}$ and finally proves the strong bulk-character of both features.

The different character of both features can be unambiguously resolved by extracting the depth dependent real part of the initial state wave function from our calculation.
The real part of the wave function $\psi(I)$ of feature (I) is shown as a purple line in Fig.~\ref{fig:figure4}b), the one of feature (II) as a blue line. 
$\psi(I)$ closely resembles the shape of a Bloch wave penetrating far into the bulk of the material (large number of layers) with (almost) constant amplitude.
The marginal decrease of the amplitude of $\psi(I)$ can be attributed to the very small but finite life times that had to be considered in our calculations.
Accordingly, feature (I) can be attributed to an optical transition between bulk states, which we will identify as the so-called Mahan cone transition\cite{Mahan.1970b,StefanHufner.2003,Winkelmann.2012}. 
In contrast, the amplitude of the wave function $\psi(II)$ of feature (II) is largest in the region close to the surface and decays towards the interior of the Ag crystal.
Crucially, the amplitude of $\psi(II)$ does not decay within a few layers to zero as expected for a Shockley like surface state (see Fig.~1), but still exhibits a non-vanishing magnitude after $600$ silver layers.
This spatial distribution of $\psi(II)$ reveals all characteristic signatures of a surface resonance.
In particular, the large amplitude of $\psi(II)$ inside the Ag bulk is responsible for the extraordinary large bulk character of this surface resonance.

After unambiguously assigning feature (I) and (II) to a bulk transition and a surface resonance with dominated bulk-contribution, we now provide a semi-quantitative explanation of the microscopic origin of both spectroscopic features.
For this, we turn to the three-step model of photoemission and construct both transitions in 3D momentum space using constant energy surfaces for the initial and final states.
So far, Mahan cone transitions have only been observed for the noble metal surfaces cut in the (111)-direction.
\begin{figure}[t]
\includegraphics{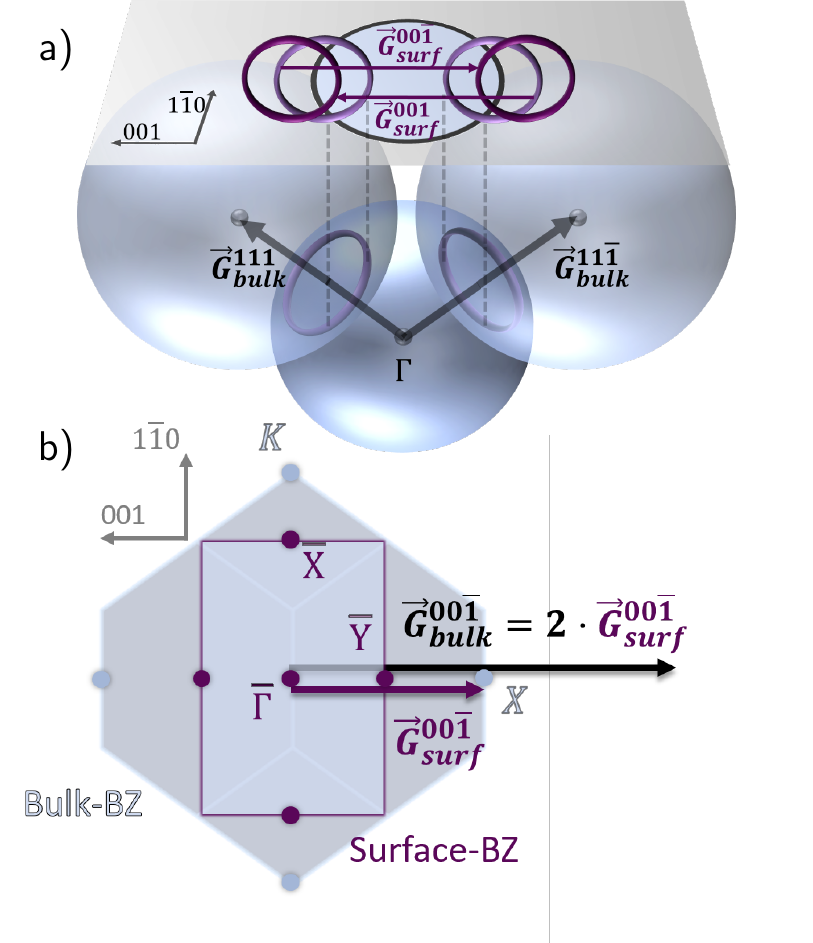}
\caption{a) Mahan cone construction in 3D momentum space based on the nearly-free electron approximation. The Mahan cone of the bulk (111)-direction is formed and projected onto the (110)-surface resulting in feature (I). Feature (II) can be constructed by moving feature (I) by a surface reciprocal lattice vector $\vec{G}_\parallel$. b) The Brillouin zone of the fcc bulk projected onto the (110)-surface and its respective surface Brillouin zone. The surface Brillouin zone provides additional reciprocal lattice vectors differing in length from their bulk counter parts.}
\label{fig:figure5}
\end{figure}
In the following, we will show that these transitions can also be observed for fcc(110) surfaces. 
We employ the nearly-free electron approximation and present the initial and final state of the Mahan cone transition as constant energy spheres in Fig.~\ref{fig:figure5}a).
Their respective radii are determined by the total momentum of the electron, $k = \sqrt{\frac{2m_{eff}E}{\hbar^2}}$ with the effective mass $m_{eff}$. The positions of the spheres of the initial and final state are given by the momentum conservation $\vec{k}_{final} = \vec{k}_{initial} + \vec{G}$, where $\vec{G}$ is an arbitrary lattice vector of the reciprocal crystal lattice.
Due to energy and momentum conservation, photoelectrons at constant energy can only be excited at the intersection of the initial and final state spheres in this momentum space construction which is highlighted by the purple circles in Fig.~\ref{fig:figure5}a).
By projecting these circles onto the Ag(110) surface, they form the parabolic features (bright purple) in the accessible field of view (grey circle) that reproduce the experimentally observed feature (I).
We can hence conclude that feature (I) is the Mahan cone transition along the $\Gamma$L-direction of bulk silver, projected onto the (110)-surface plane.
Additionally, we included a construction based on our LDA calculations in the supporting information.

The surface resonance with large bulk character can be explained by additional contributions of surface associated reciprocal lattice vectors $\vec{G}_\parallel$ to the Bloch states of bulk electrons of the fcc lattice.
These additional $\vec{G}_\parallel$ arise due to the mismatch between the surface projected Brillouin zone of the fcc bulk crystal and the surface Brillouin zone of the fcc(110) surface illustrated in Fig.~\ref{fig:figure5}b).
The momentum conversation transforms into $\vec{k}_{final} = \vec{k}_{initial} + \vec{G}_\parallel+ \vec{G}$.
This surface-induced modification of the momentum conservation allows us to consider another translation of the surface projected Mahan Cone feature by exactly one reciprocal surface lattice vectors, in this case $\vec{G}_\parallel = (0,0,\pm1)\cdot \frac{2\pi}{a}$, where $a$ is the surface lattice constant.
This results in a new state located in the surface following the momentum dispersion of the bulk state, which we will refer to as \emph{surface scattering resonance}.
Crucially, this surface scattering resonance is not simply a replica of the bulk Mahan Cone scattering at the surface as it reveals its own light polarization and photon energy dependent photocurrent.
We hence propose that the surface scattering resonance is truly a new type of surface resonance, which arises from additional surface contributions to the Fourier sum of the bulk Bloch states depending on $\vec{G}_\parallel$.

Based on our assignment of features (I) and (II) to the Mahan cone and the surface scattering resonance, we propose an empirical rule for the polarization dependent emission pattern of both features discussed in Fig.~\ref{fig:figure3}.
The measured photoemission signal $S(\vec{k})$ can be reproduced by including two simple rules 
\begin{itemize}
	\item[(I)] $S(\vec{k}) \propto 1 - \cos\alpha$	
	\item[(II)] $S(\vec{k}) \propto \cos\alpha$
\end{itemize}
regarding the angle $\alpha$ between $\vec{k}$ and $\vec{E}$ in the geometrical construction, with $\vec{k}$ being constrained within the surface layer for feature (II).
This phenomenological model further support our identification of both features as the photoemission yield of the Mahan Cone can be explained by the orientation of the light polarization with respect to the 3D Brillouin zone while the one of the surface scattering resonance is only sensitive to the relative orientation of the light polarization vector and the crystal momentum of the 2D surface Brillouin zone.

In conclusion, our work uncovers clear spectroscopic evidence for a new type of surface resonance with exceptionally large bulk contribution in the surface band structure of an Ag(110) single crystal.
This so called surface scattering resonance arises due to a lattice mismatch between the surface projected bulk Brillouin zone and the surface Brillouin zone of the Ag(110) surface leading to additional surface-related contributions to the bulk Bloch states of bulk transitions. 
We propose that these types of surface resonances can be observed for a vast majority of surfaces since similar lattices mismatches between surface projected bulk Brillouin zone and the surface Brillouin zone are the rule rather than the exception for most single crystal surfaces.
This will potentially set the stage to discover new, possibly exotic, surface resonances with exceptional bulk character in a huge variety of quantum or topological materials.

\begin{acknowledgements}
The research leading to these results was funded by the Deutsche Forschungsgemeinschaft (DFG, German Research Foundation) – TRR 173 – 268565370 (project A02) and EB154/37. 
\end{acknowledgements}

\end{document}

\title{Spectroscopic evidence for a new type of surface resonance at noble metal surfaces \\ Supplemental Material}
\author{Tobias Eul}
\email[]{teul@rhrk.uni-kl.de}
\affiliation{Technische Universität Kaiserslautern and Research Center OPTIMAS, 67663 Kaiserslautern, Germany}
\author{Jürgen Braun}
\affiliation{Department Chemie, Ludwig-Maximilians-Universität München, 81377 München, Germany}
\author{Benjamin Stadtmüller}
\affiliation{Technische Universität Kaiserslautern and Research Center OPTIMAS, 67663 Kaiserslautern, Germany}
\author{Hubert Ebert}
\affiliation{Department Chemie, Ludwig-Maximilians-Universität München, 81377 München, Germany}
\author{Martin Aeschlimann}
\affiliation{Technische Universität Kaiserslautern and Research Center OPTIMAS, 67663 Kaiserslautern, Germany}
%\date{\today}

\maketitle

\begin{figure}[!b]
\includegraphics{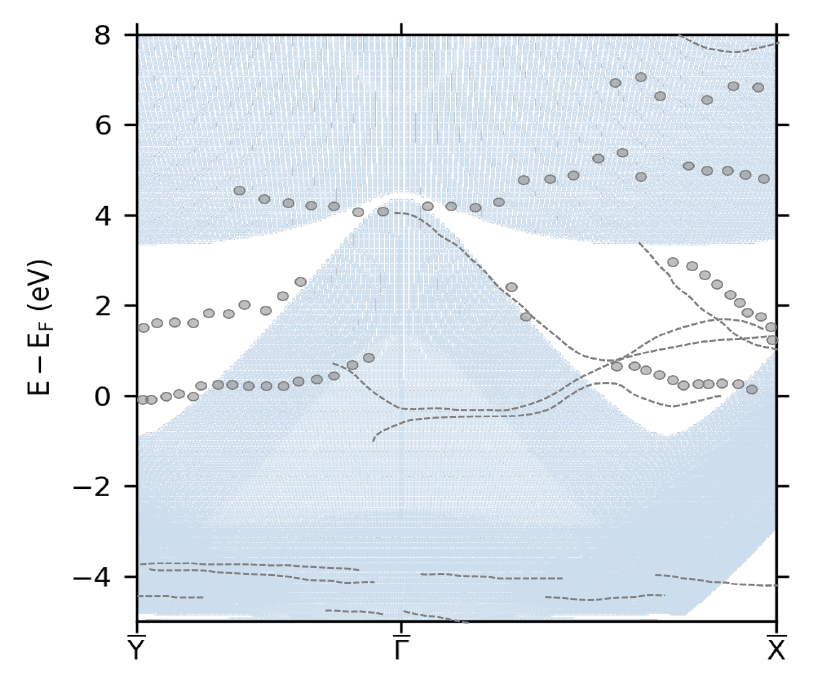}%
\caption{Calculated bulk projected band structure of Ag(110) in light blue. Previously reported experimental (dots) and theoretical (dashed) data from\cite{Ho.1980,Altmann.1986} is depicted in grey for comparison.}
\label{fig:figure0}
\end{figure}

\section{Sample preparation}
The Ag(110) single crystal was prepared by Ar$^+$-sputtering and subsequent annealing.
The Argon gas (pressure $\approx \SI{6e-6}{\milli\bar}$) was ionized with a current of \SI{10}{\milli\ampere} and accelerated towards the sample with a voltage of \SI{1500}{\volt}.
The surface normal of the sample was aligned with the optical axis of the sputter gun for normal incidence sputtering.
The sample was subsequently heated to \SI{550}{\celsius}.
The duration of a sputtering cycle was typically in the range of \SIrange{20}{30}{\minute} and the following annealing cycle twice the duration of the sputtering.

\section{Calculation details}
For a realistic description of the surface electronic structure, we additionally included a spin-independent Rundgren-Malmström potential \cite{G.Malmstrom.1980}, which represents a one-dimensional z-dependent function $V_{B}(z)$ with the z-axes directed perpendicular to the surface pointing into the semi-infinite bulk to guarantee a proper asymptotic behaviour of the potential. 
The potential $V_{B}(z)$ connects the asymptotic regime ($V_{B}(z) \propto 1/z$,\,$z_{A}>z$ ) to the bulk muffin-tin zero $V_{\textrm or}$ by a third order polynomial $P(z)$, e.g. defining a polynomial range $z_{E}>z>z_{A}$. 
The second parameter $z_{E}$ defines the point where the surface region ends and the bulk region starts. 
The third parameter $z_{I}$ ($z_{E}>z_I>z_{A}$), which allows to modify the shape of the potential function, is typically identified with the position of the classical image plane \cite{G.Malmstrom.1980}. 
We have calibrated these three parameters in a way that the calculations reproduce the experimental band dispersion of the existing surface resonances on Ag(110)\cite{L.E.Urbach.1992,A.Gerlach.1999,Pascual.2001}.
As described in the main text, we alter the lifetimes of the initial and final states to control the bulk or surface sensitivity.
To simulate an artificially increased bulk sensitivity, we phenomenologically attribute the line width of the final state a constant imaginary value of $V_i$ = \SI{0.1}{\electronvolt}, while we set the hole life-time of the initial state to a constant imaginary value of $V_i$ = \SI{0.12}{\electronvolt}. 

\section{Polarization dependent photoemission yield}
\begin{figure}[hb]
\includegraphics{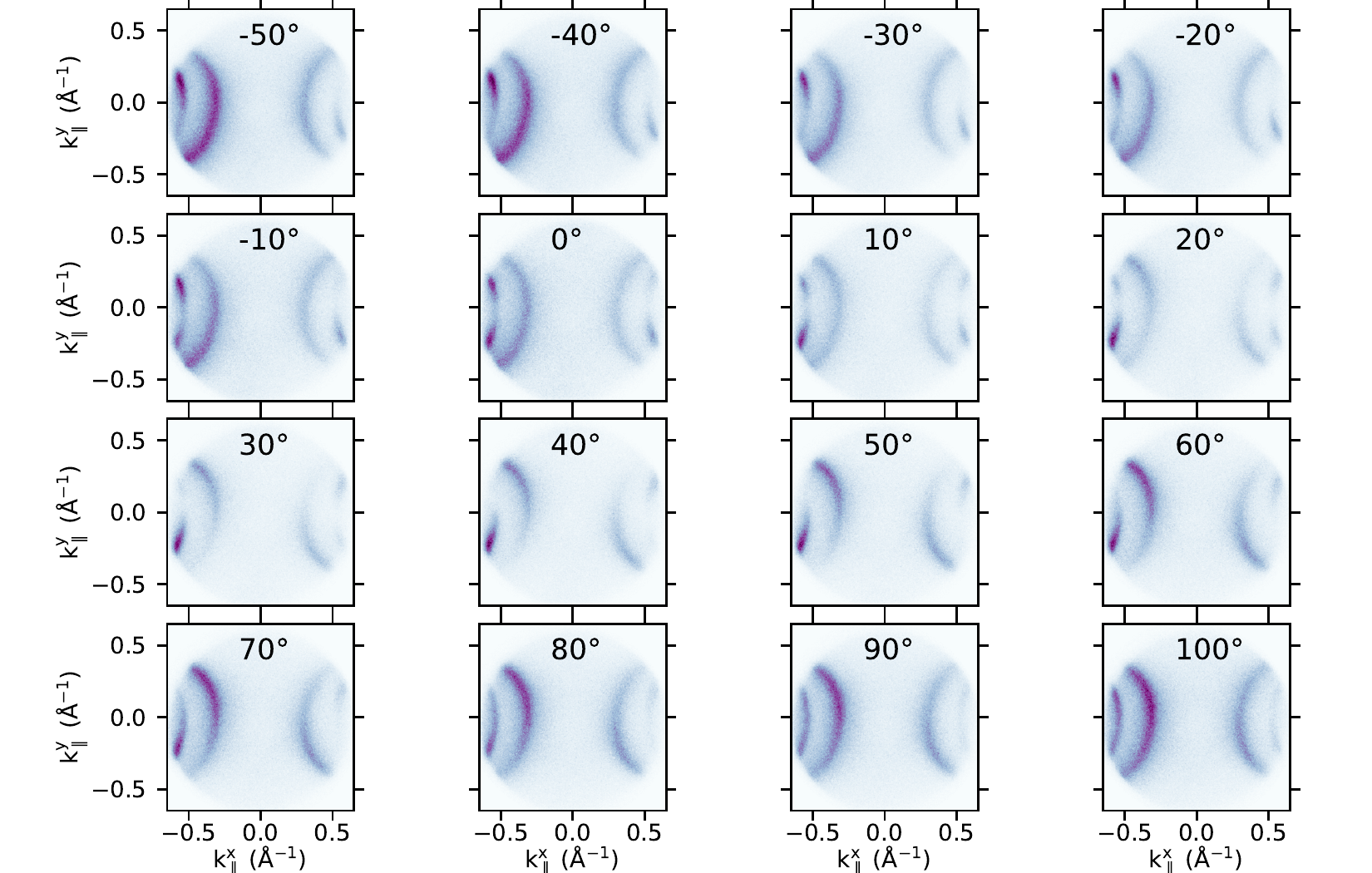}
\caption{Constant binding energy maps for photoemission in the normal incidence excitation geometry for a photon energy of \SI{5.9}{\electronvolt} with varying angle $\varphi$ between the electric field and the $\overline{\Gamma}\,\overline{\textrm{Y}}$-axis. The images were taken at $E-E_F=\SI{0}{\electronvolt}$. The images have been rotated in order to align the $\overline{\Gamma}\,\overline{\textrm{Y}}$-direction along the $k_\parallel^x$-axis. $\varphi$ has been varied from \SIrange{-50}{100}{\degree} in steps of \SI{10}{\degree}.}
\label{fig:figure1}
\end{figure}

In this section we present additional constant binding energy maps corresponding to those shown in Fig. 3 in the main manuscript.
Figure \ref{fig:figure1} shows CBMs for additional angles $\varphi$ from \SIrange{-50}{100}{\degree} in steps of \SI{10}{\degree} between the light polarization vector and the $\overline{\Gamma}\overline{\textrm{Y}}$-direction of the surface Brillouin zone in normal incidence geometry.
\clearpage

\section{Geometrical Mahan Cone construction}
In this section we present the details of the geometrical construction of the Mahan Cone based on the free electron model for the final state.
In the past, the Mahan cone has been constructed geometrically based on a free-electron model in 3D using spherical constant energy surfaces to represent experimental data qualitatively.
This construction is based on the assumption that an electron can only be excited  from the initial to a final state by fulfilling the condition $\vec{G}_{111} = \vec{k}_{i} + \vec{k}_{f}$, where $\vec{G}_{111}$ is a reciprocal lattice vector.
Geometrically, this is represented by two spheres with radii $|\vec{k}_{f}|$ and $|\vec{k}_{i}|$ positioned at $\Gamma$ and $\Gamma'$ respectively\cite{Mahan.1970b,StefanHufner.2003,Winkelmann.2012}.
The emission condition is then fulfilled for all momentum vectors pointing to the intersection of the spheres.
To find the free electron parameters for the construction of the Mahan cone, we reproduced photoemission data from a Ag(111)-surface taken with a photon energy of \SI{5.9}{\electronvolt}.
We extract the work function as $\phi=\SI{4.3}{\electronvolt}$ from the measurement.
Next, we identified the maximum kinetic energy of the electron in the sp-band as $E_{kin}=-E_0=\SI{4.9}{\electronvolt}$ by comparing with bulk band structure calculations.
Together, they yield the inner potential for the construction $V_0 = \SI{9.2}{\electronvolt}$ and the final parameter to be adapted is the effective mass of $m^\ast=0.63m_e$.

\begin{figure}[t]
\includegraphics[width=0.5\textwidth]{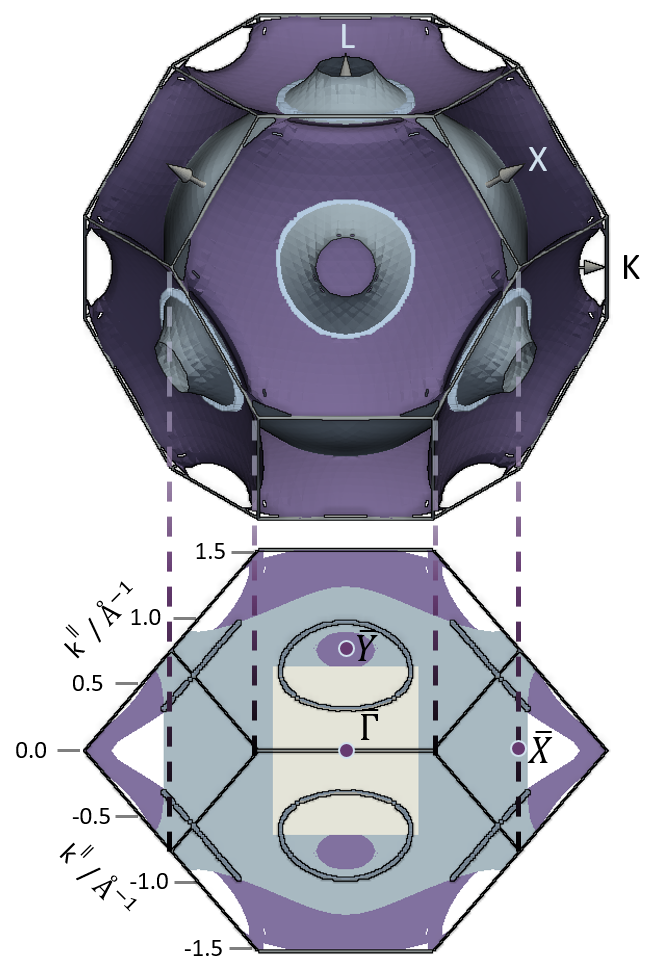}
\caption{Mahan cone construction in the 3D Brillouin zone of a fcc crystal. Calculated Fermi surface of Ag in grey and the possible final states for a photonenergy of \SI{5.9}{\electronvolt} in purple, respectively. The glowing grey line shows the intersection of the two surfaces which result in the Mahan cone when including all possible initial and final states. The projection of the 3D Brillouin zone onto the 110-surface resulting in the parabolic feature in the experiment is depicted at the bottom. The bright rectangle reflects the accessible momentum range in the experiment.}
\label{fig:figure2}
\end{figure}

Addtionally, we employed our self-consistent LDA calculations to realize the constant energy surfaces of the initial and final state, which more accurately describes the experiment than the more common approach of the free-electron model.
These constant energy surfaces and the corresponding Mahan cone construction in the first Brillouin zone is shown in  Fig.~\ref{fig:figure2}a) for our experiment on the Ag(110) crystal.
The image displays the three-dimensional Brillouin zone of the fcc crystal viewed along the (111)-direction.
Within the Brillouin zone, the Fermi surface of Ag is displayed in grey representing the initial state of the photoemission process.
The purple planes represent constant energy surfaces of the final states using a photon energy of \SI{5.9}{\electronvolt}.

This is highlighted by the grey line around the L-point, representing a cut through the Mahan cone of the (111)-direction for electrons excited at $E-E_F$=\SI{0}{\electronvolt}.
The bottom part of Fig.~\ref{fig:figure2} displays the projection of the 3D-Brillouin zone onto the (110)-surface representing our experimental situation.
Due to the projection, we see circular intersection lines in both the positive and negative $\overline{\Gamma}\,\overline{\textrm{Y}}$-direction.

\section{Additionally extracted wave functions}

\begin{figure}[t]
\includegraphics{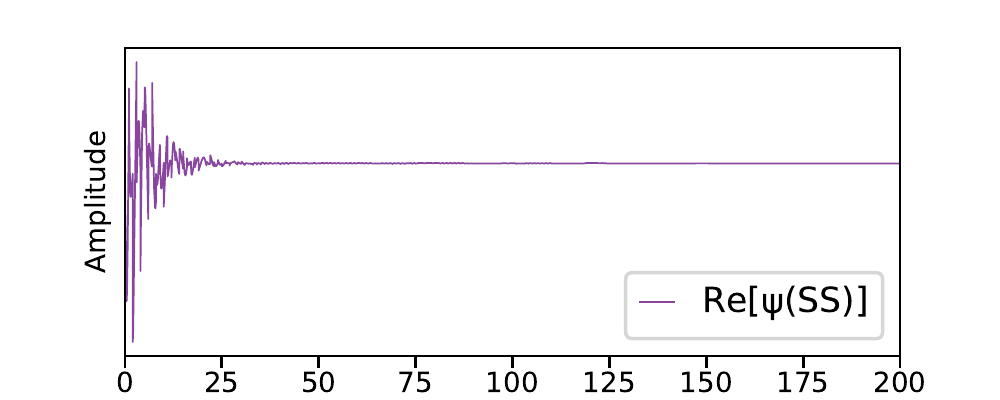}
\caption{Real part of the wave function extracted from the surface state located in the bandgap around the $\overline{Y}$-point of the surface Brillouin zone. The amplitude decays already in the first surface layers.}
\label{fig:figure3}
\end{figure}

In addition to the wave functions already displayed in the main manuscript, we extracted the real part of the wave function from the surface state located in the band gap around the $\overline{Y}$-point of the surface Brillouin zone.
This state has been previously reported, for example in\cite{L.E.Urbach.1992,A.Gerlach.1999}.
The surface state became accessible with our photocurrent calculations with increased photon energies.
This wave function was extracted from the calculations for $h\nu=\SI{8}{\electronvolt}$ and is shown in Fig.~\ref{fig:figure3}.
Here, its amplitude already decays within the first few layers of the sample, contrary to the wave functions of the Mahan Cone feature and the surface scattering resonance which both propagate deep into the interior of the sample.

\clearpage